# Subband anticrossing and the spin Hall effect in quantum wires

A.W. Cummings, R. Akis, and D.K. Ferry

Department of Electrical Engineering and Center for Solid State Electronics Research, Arizona State University, Tempe, Arizona 85287-5706

We report on numerical simulations of the intrinsic spin Hall effect in semiconductor quantum wires as a function of the Rashba spin-orbit coupling strength, the electron density, and the width of the wire. We find that the strength of the spin Hall effect does not depend monotonically on these parameters, but instead exhibits a local maximum. This behavior is explained by considering the dispersion relation of the electrons in the wire, which is characterized by the anticrossing of adjacent subbands. These results lead to a simple estimate of the optimal wire width for spin Hall transport experiments, and simulations indicate that this optimal width is independent of disorder. The anticrossing of adjacent subbands is related to a quantum phase transition in momentum space, and is accompanied by an enhancement of the Berry curvature and subsequently in the magnitude of the spin Hall effect.

In recent years, a great deal of attention has been given to spin-orbit coupling as a means of providing electronic control over the spin of electrons in nanoelectronic devices. Of particular interest is Rashba spin-orbit coupling [1], which is present in a two-dimensional electron gas (2DEG) that lies in an asymmetric quantum well. The Hamiltonian describing the Rashba spin-orbit interaction in a 2DEG is given by

$$H_R = \alpha_z \left( \sigma_x k_y - \sigma_y k_x \right), \tag{1}$$

where $\alpha_z$ represents the strength of the interaction, $\sigma_i$ represents the Pauli matrices, and $k_i$ are the in-plane components of the wave vector of the electron. The *z*-axis is taken to be along the growth direction of the heterostructure, while the 2DEG sits in the *xy*-plane. Because of its dependence on the shape of the quantum well and the nature of the wave function along the growth direction, some measure of control over the parameter $\alpha_z$ is possible with proper design of the heterostructure, and finer tuning can be accomplished with an electrostatic top gate [2]. Rashba spin-orbit coupling is characterized by a *k*-linear splitting of the two-dimensional band structure, where the spin-split subbands correspond to electrons whose spin is oriented in the plane of the 2DEG and perpendicular to the direction of transport [3]. The lifting of the spin degeneracy results in two different carrier densities within the 2DEG, and the resulting beating pattern in Shubnikov-de Haas measurements can be used to determine the magnitude of $\alpha_z$ [4].

Another consequence of Rashba spin-orbit coupling is known as the intrinsic spin Hall effect, where a longitudinal charge current is accompanied by a transverse spin current, polarized normal to the plane of the 2DEG [5]. In finite systems such as quantum wires, the transverse spin current leads to an accumulation of oppositely polarized spins on opposite sides of the wire [6]; this has led to several proposals of Y-shaped branching structures as a means of generating spin-polarized currents in mesoscopic systems [7-11]. Most experimental efforts to measure the spin Hall effect in semiconductor systems have focused on optical techniques [12,13]. However, recent work of ours has shown that it is possible to use a double Y-branch structure, in conjunction with the spin Hall effect, to generate and detect spin-polarized currents in InAs quantum wells in a purely electrical measurement [14,15].

In a typical mesoscopic experiment, a quantum wire is formed by confining the 2DEG in a semiconductor heterostructure along a particular axis. This can be accomplished with electrostatic top gates or with a lithographic etching process. While our previous work has focused on electron transport through Y-junction nanowire structures in order to generate and detect spin-polarized currents [7,14,15], the goal of this paper is to more closely examine the behavior of the spin Hall effect in straight quantum wires, with an eye toward experimental optimization. Therefore, we study the dependence of the spin Hall effect in quantum wires on three fundamental parameters - the electron density in the 2DEG formed by the heterostructure, the Rashba spin-orbit coupling strength, and the width of the wire. To do this, we start with the Hamiltonian for an electron in a 2DEG, $H = H_0 + H_R$. $H_R$ is given in (1) and $H_0$ is given as

$$H_0 = \frac{\hbar^2}{2m^*}\left(k_x^2 + k_y^2\right) + V(x,y), \tag{2}$$

where $V(x,y) = 0$ for $0 \leq x \leq W$ and is infinite otherwise. Next, we apply the Hamiltonian to the time-independent Schrödinger equation, $H\psi = E\psi$, and assume a spin-resolved form of the wave function,

$$\psi = e^{iky}\begin{bmatrix}\phi_\uparrow(x)\\ \phi_\downarrow(x)\end{bmatrix}. \tag{3}$$

Here we have chosen the *y*-axis to be the direction of propagation, and can assume a plane wave form due to the translational invariance of the Hamiltonian along this axis. The functions $\phi_\uparrow(x)$ and $\phi_\downarrow(x)$ represent the spin-up and spin-down wave functions along the confinement axis of the wire, where spin-up is oriented along +*z* and spin-down along –*z*. For the sake of simplicity, from this point forward we drop the explicit dependence of these functions on *x*, but this dependence is still implied. By inserting the 2 x 2 form of the Pauli matrices, substituting the operators $k_u = -i\dfrac{\partial}{\partial u}$, and applying (3) to the Schrödinger equation, we arrive at a pair of coupled equations for $\phi_\uparrow$ and $\phi_\downarrow$,

$$\phi_\uparrow'' - k^2\phi_\uparrow - 2k_{SO}\phi_\downarrow' - 2kk_{SO}\phi_\downarrow = -\dfrac{2mE}{\hbar^2}\phi_\uparrow, \tag{4a}$$

$$\phi_\downarrow'' - k^2\phi_\downarrow + 2k_{SO}\phi_\uparrow' - 2kk_{SO}\phi_\uparrow = -\dfrac{2mE}{\hbar^2}\phi_\downarrow, \tag{4b}$$

where $k_{SO} = \dfrac{m^*\alpha_z}{\hbar^2}$. In order to solve these equations numerically, we discretize the $x$-axis such that (4a) and (4b) become

$$\dfrac{1}{a_x^2}\left(\phi_\uparrow^{m+1} + \phi_\uparrow^{m-1}\right) - \left(\dfrac{2}{a_x^2} + k^2\right)\phi_\uparrow^m - \dfrac{k_{SO}}{a_x}\left(\phi_\downarrow^{m+1} - \phi_\downarrow^{m-1}\right) - 2kk_{SO}\phi_\downarrow^m = -\dfrac{2mE}{\hbar^2}\phi_\uparrow^m, \tag{5a}$$

$$\dfrac{1}{a_x^2}\left(\phi_\downarrow^{m+1} + \phi_\downarrow^{m-1}\right) - \left(\dfrac{2}{a_x^2} + k^2\right)\phi_\downarrow^m + \dfrac{k_{SO}}{a_x}\left(\phi_\uparrow^{m+1} - \phi_\uparrow^{m-1}\right) - 2kk_{SO}\phi_\uparrow^m = -\dfrac{2mE}{\hbar^2}\phi_\downarrow^m, \tag{5b}$$

where $\phi_{\uparrow,\downarrow}^m$ represents the value of the confined wave function at grid point $m$, and $a_x$ is the distance between grid points. Finally, (5a) and (5b) can be combined into a single eigenvalue equation,

$$\begin{bmatrix} H_{11} & H_{12} \\ H_{21} & H_{22} \end{bmatrix}\begin{bmatrix} \phi_\uparrow \\ \phi_\downarrow \end{bmatrix} = \dfrac{-2m^*E}{\hbar^2}\begin{bmatrix} \phi_\uparrow \\ \phi_\downarrow \end{bmatrix}, \tag{6a}$$

$$H_{11} = H_{22} = -\dfrac{1}{a_x^2}\begin{bmatrix} 2+a_x^2k^2 & -1 & & \\ -1 & \ddots & \ddots & \\ & \ddots & \ddots & -1 \\ & & -1 & 2+a_x^2k^2 \end{bmatrix}, \tag{6b}$$

$$\text{and } H_{12} = H_{21}^T = -\dfrac{k_{SO}}{a_x}\begin{bmatrix} 2a_xk & 1 & & \\ -1 & \ddots & \ddots & \\ & \ddots & \ddots & 1 \\ & & -1 & 2a_xk \end{bmatrix}. \tag{6c}$$

The matrices $H_{ij}$ all have dimension $n \times n$, where $n$ is the number of grid points along the confinement axis. Equations (6a-c) can be used to find the subband energies and their corresponding wave functions for a given value of $k$.

In a two-dimensional system, the spin Hall conductivity can be calculated by replacing the transverse charge current operator with the transverse spin current operator in the Kubo formula [5]. Shi *et al.* [16] have defined the spin current operator as the time derivative of the spin displacement operator, $\hat{\mathbf{J}}_S = d(\mathbf{r}s_z)/dt$. However, when applied along the *x*-axis, this operator vanishes in our system because $\phi = [\phi_\uparrow \quad \phi_\downarrow]^T$ is a localized eigenstate and thus has no time dependence. Therefore, we choose to characterize the strength of the spin Hall effect in a particular subband by using the spin displacement operator directly,

$$\langle x\sigma_z \rangle = \int_x \phi^*(x\sigma_z)\phi dx = \int_x \phi_\uparrow^* x\phi_\uparrow dx - \int_x \phi_\downarrow^* x\phi_\downarrow dx = \langle x_\uparrow \rangle - \langle x_\downarrow \rangle. \tag{7}$$

The spin displacement operator allows a direct comparison of the strength of the spin Hall effect to the width of the quantum wire. If we assume the low-bias regime, low temperature, and consider electron transport in the $+y$ direction, then we can limit ourselves to an investigation of the subbands at the point where they cross the Fermi energy with a positive slope, $dE/dk$. To find the total spin displacement due to the modes that contribute to the forward conductance, each subband that crosses the Fermi energy is weighted by its 2D density of states, $(k - k_{0n})/(2\pi \cdot dE/dk)$, where $k_{0n}$ is the wave number at the minimum of the *n*th subband.

In figure 1 we plot the spin displacement as a function of the wire width for different values of $\alpha_z$, assuming an InAs quantum well with an electron density of 4 x $10^{11}$ cm$^{-2}$. The vertical axis has been normalized such that it corresponds to the spin displacement as a fraction of the wire width, $\langle x\sigma_z \rangle/W$. Figure 1(a) shows the spin displacement when the contribution of all the occupied subbands is considered. In this figure we see the general trend that $\langle x\sigma_z \rangle/W$ increases with the spin-orbit coupling strength. The oscillations of $\langle x\sigma_z \rangle/W$ correspond to the population of higher subbands as the wire width increases. We also see that $\langle x\sigma_z \rangle/W$ at first increases with the wire width, but then levels off and starts to decrease. This trend is more evident in figure 1(b), where $\langle x\sigma_z \rangle/W$ has been calculated considering only the lowest occupied pair of subbands. Here, the spin displacement increases with wire width, reaches a maximum, and then decreases and eventually reverses sign. The width at which $\langle x\sigma_z \rangle/W$ is maximized varies inversely with the spin-orbit strength. This is reminiscent of the behavior seen by Moca and Marinescu [17], who calculated an optimal wire width on the order of the spin precession length in the presence of disorder. Similar size dependence can be seen in figure 2, where we plot $\langle x\sigma_z \rangle/W$ as a function of wire width for different values of the electron density, assuming a spin-orbit coupling strength of $\alpha_z$ = 20 meV-nm. In figure 2(a), where we plot the effect of all occupied subbands, we see that the spin displacement decreases with increasing electron density, due to the interference of multiple occupied subbands. In figure 2(b), where the spin displacement due to only the lowest pair of subbands is plotted, the width where $\langle x\sigma_z \rangle/W$ is maximized varies inversely with electron density.

To explain the results seen in figures 1 and 2, we first consider figure 3(a), which shows the dispersion relation of a 100-nm InAs wire with a spin-orbit strength of $\alpha_z$ = 20 meV-nm. The energy axis has been scaled by the bare kinetic energy term, such that $E_{SO}(k) = E(k) - \frac{\hbar^2 k^2}{2m^*}$. The results in figure 3(a) are similar to those found by Moroz and Barnes [18], where the *k*-linear splitting of the subbands holds for small values of *k*, while at higher values of *k* the dispersion relation is characterized by the anticrossing of adjacent subbands. As discussed by Moroz and Barnes, anticrossing corresponds to the hybridization of adjacent subbands. In figure 3(b), we show the spin displacement of each subband as a function of *k*, corresponding to the plot in figure 3(a). There are several interesting features present in this figure. First, we see that the spin displacement of subband **1-** decays as $1/k$. While the Rashba Hamiltonian is linear in $k$, the bare Hamiltonian representing the kinetic energy is proportional to $k^2$. Therefore, for large values of $k$, the wave function corresponding to the lowest band will be dominated by the bare solution, when $\alpha_z = 0$. Second, we see that the extrema in the spin displacement of the higher subbands correspond to the anticrossing points in figure 3(a). For larger values of $k$, the spin displacement then decays, similar to subband **1-**. Higher-order anticrossings, such as the one between subbands **2-** and **3+** at $k = 0.6$ nm$^{-1}$, also exhibit a small enhancement of the spin displacement, although not nearly to the degree of the first anticrossing point, due to the dominance of bare Hamiltonian.

By considering figure 3, one can explain the notable features of figures 1 and 2. As the width of the wire is increased, the subband energies move downward through the Fermi energy. Whenever subband **n+** coincides with the Fermi energy at its anticrossing point, the spin separation exhibits a local maximum. This explains both the presence of the maxima in figures 1(b) and 2(b), and the oscillatory behavior of the spin displacement in figures 1(a) and 2(a). We also note that in figures 1(b) and 2(b), the spin displacement actually changes sign when $W$ becomes large enough. For large values of $W$, subbands **1-** and **1+** cross the Fermi energy at a relatively large value of $k$, well beyond the anticrossing point of **1+**. At this point, the spin displacements of these two subbands are approximately equal, but opposite in sign. However, subband **1-** has a higher density of states, and thus the total spin displacement becomes negative in this region.

The above results suggest a couple methods of optimizing the spin Hall effect in a mesoscopic transport experiment. A comparison of figures 1(a) and 2(a) with figures 1(b) and 2(b) shows that the spin Hall effect is significantly enhanced when only the lowest pair of subbands is occupied. One way to achieve this situation in a quantum wire is to use a quantum point contact (QPC) to filter out the higher subbands, so that only the lowest pair of subbands pass through the QPC. To optimize the spin displacement in this situation, one can measure the spin-orbit strength and the electron density of the 2DEG, and use that information to choose a wire width such that the Fermi energy coincides with the anticrossing point of subbands **1+** and **2-**; $E_1^+(k_F) = E_2^-(k_F)$, where $k_F$ is the Fermi wave number. If we assume that an etched quantum wire can be approximated by an infinite square well potential, then the subband energies in the linear splitting regime are given by $E_n^\pm(k) = \frac{\hbar^2}{2m^*}\left(\frac{\pi^2 n^2}{W^2} - k_{SO}^2 + k^2\right) \pm \alpha_z k$ [19] and the wire width that maximizes the spin Hall effect after filtering by a QPC is given by

$$W = \sqrt{\frac{3\pi^2 \hbar^2}{4m^* \alpha_z (2\pi n_{2D})^{1/2}}}. \tag{8}$$

Assuming $\alpha_z = 20$ meV-nm and $n_{2D} = 4 \times 10^{11}$ cm$^{-2}$, we have $W = 88$ nm, corresponding to what we see in figures 1(b) and 2(b).

Instead of using a QPC, the application of a top- or back-gate voltage can reduce the density of electrons, and subsequently the Fermi energy, in the wire until only the lowest pair of subbands is occupied. However, it will be important to design the heterostructure such that the application of a gate bias does not effect the spin-orbit strength [20,21]. In this case, to maximize the spin displacement we want the anticrossing point of subbands **1+** and **2-** to occur at an energy lower than the energy of subband **2-** at $k = 0$. Using the expression for the subband energies given above, this condition is satisfied when

$$W > \frac{\sqrt{3}\pi \hbar^2}{4m^* \alpha_z}. \tag{9}$$

Using $\alpha_z = 20$ meV-nm, we find that $W > 225$ nm in this case.

Disorder can also play a significant role in the manifestation of the spin Hall effect in semiconductor heterostructures. In the infinite 2D system, it was found that arbitrarily weak isotropic scattering produces a vertex correction that exactly cancels out the spin Hall conductivity [22]. However, subsequent studies have shown that the spin Hall effect persists in finite 2D systems in the presence of disorder [23,24]. To study the effect of disorder in a quantum wire, we can no longer restrict ourselves to a particular slice in the wire, because the disorder potential eliminates the translational invariance of the Hamiltonian. Instead, we use a two-dimensional cascaded scattering matrix approach discussed by Usuki *et al.* [25] which has been augmented with the Rashba Hamiltonian [14]. The disorder potential is modeled as a random energy at each lattice site, and the energies are subject to a Lorentzian probability density function, $P(E) = \frac{1}{\pi} \frac{\Gamma}{\Gamma^2 + E^2}$, where $\Gamma$ is the half-width at half maximum [26]. We assume a wire with a length of 1 μm, and for each instance of the disorder potential we calculate an average spin displacement over the length of the wire. We then average this over 500 instances of the disorder potential, and to get an idea of the variability of the results, we also calculate the standard deviation of the spin displacement. The results can be seen in figure 4, where we plot the spin displacement for $\Gamma$ = 1, 3, and 5 meV, assuming an electron density of $n_{2D}$ = 4 x $10^{11}$ cm$^{-2}$ and a spin-orbit strength of $\alpha_z$ = 20 meV-nm, considering only the lowest pair of occupied subbands. The error bars indicate the standard deviation of the results. To get an idea of the energy scales and the relative magnitude of $\Gamma$, we note that at the point of anticrossing between subbands **1+** and **2-**, $E_1^+ - E_1^-$ = 4.3 meV and $E_2^- - E_1^+$ = 1.06 meV. In figure 4, we see that the spin Hall effect is suppressed but not completely eliminated by the presence of disorder in the wire, and that the optimal wire width remains unchanged.

Finally, it is important to note that we can explain the peaks in the spin separation that appear in figures 1-3 by considering the role of the topological (geometric) phase that occurs in spin transport. A great deal of attention has been given to the geometric phase in 2D systems and its connection to the quantum Hall effect [27,28], the anomalous Hall effect [29,30], and the spin Hall effect [31]. In particular, it has been shown that the geometric, or Berry's [32], phase is directly proportional to the various types of Hall conductivities in these systems. To develop the concept of geometric phase, we follow in the footsteps of Resta [33], and start with a generic quantum mechanical system described by $H(\mathbf{R})|n(\mathbf{R})\rangle = E_n(\mathbf{R})|n(\mathbf{R})\rangle$, where the Hamiltonian and its eigenstates depend on a set of parameters represented by $\mathbf{R}$. The overlap integral of an eigenstate at two different points in parameter space is given by $\langle n(\mathbf{R}_1)|n(\mathbf{R}_2)\rangle = |\langle n(\mathbf{R}_1)|n(\mathbf{R}_2)\rangle| \cdot e^{i\Delta\varphi_{12}}$. In addition to the magnitude term, the overlap integral acquires a phase term, $\Delta\varphi_{12} = \text{Im}(\ln\langle n(\mathbf{R}_1)|n(\mathbf{R}_2)\rangle)$, which represents the difference in local phase between an eigenstate located at $\mathbf{R}_1$ and $\mathbf{R}_2$. However, the phase of a particular eigenstate is an arbitrary quantity, and $\Delta\varphi_{12}$ can be made to vanish by an appropriate choice of gauge at either $\mathbf{R}_1$ or $\mathbf{R}_2$. Therefore, $\Delta\varphi_{12}$ cannot represent a physical observable of the system. However, this difficulty can be overcome if we consider the phase accumulated around a complete loop in parameter space. For example, in a 3-point loop we find that

$\gamma = \Delta\varphi_{12} + \Delta\varphi_{23} + \Delta\varphi_{31} = \text{Im}(\ln\langle n(\mathbf{R}_1)|n(\mathbf{R}_2)\rangle\langle n(\mathbf{R}_2)|n(\mathbf{R}_3)\rangle\langle n(\mathbf{R}_3)|n(\mathbf{R}_1)\rangle)$. Here we see that the gauge-dependent phase terms cancel out in pairs, and the overall phase is gauge-invariant. Therefore, $\gamma$ can represent a physical observable.

The discrete geometric phase described above can be extended to the continuous case if we take the overlap between two eigenstates separated by an arbitrarily small distance in parameter space, $\Delta\mathbf{R}$. In this case, $\langle n(\mathbf{R})|n(\mathbf{R}+\Delta\mathbf{R})\rangle \approx e^{-i\Delta\mathbf{R}\cdot\mathbf{A}_n(\mathbf{R})}$, where $\mathbf{A}_n(\mathbf{R}) = i\langle n(\mathbf{R})|\nabla_\mathbf{R}|n(\mathbf{R})\rangle$ is known as the Berry connection [30]. Thus, the phase due to a loop in parameter space takes the well-known form of the Berry phase [32], $\gamma_B^n = \oint \mathbf{A}_n(\mathbf{R}) d\mathbf{R}$. Using Stokes' theorem, this can also be written as $\gamma_B^n = \iint \mathbf{\Omega}_n(\mathbf{R}) \cdot d\mathbf{S}$, where $\mathbf{\Omega}_n(\mathbf{R}) = \nabla_\mathbf{R} \times \mathbf{A}_n(\mathbf{R})$.

An early example of the connection between the Berry phase and a physical observable was illustrated by Thouless [27] and Kohmoto [28] in their study of the quantum Hall effect. In particular, they showed that the Hall conductivity of a 2DEG in a perpendicular magnetic field is proportional to the Berry phase and can be written as $\sigma_H = \frac{e^2}{h} \frac{1}{2\pi i} \sum_n \oint d\mathbf{k} \cdot \mathbf{A}_n(\mathbf{k})$. Later, Sundaram and Niu showed that the velocity operator of a semiclassical wave packet acquires an anomalous term proportional to the Berry curvature, $\Omega_n(k) = i\left(\left\langle \frac{\partial u_n}{\partial k_1} \middle| \frac{\partial u_n}{\partial k_2} \right\rangle - \left\langle \frac{\partial u_n}{\partial k_2} \middle| \frac{\partial u_n}{\partial k_1} \right\rangle\right)$, where $u_n$ is the cell-periodic part of the Bloch function [34]. Jungwirth *et al.* [29] connected this formalism to the manifestation of the anomalous Hall effect in ferromagnets, and showed that the anomalous Hall conductivity is proportional to the Berry phase. Meanwhile, Fang *et al.* [30] and Onoda *et al.* [35] highlighted the importance of band crossings and anticrossings in connection with the anomalous Hall effect. In particular, it was pointed out that the intrinsic anomalous Hall effect is resonantly enhanced when the Fermi level lies near the anticrossing points of subbands split by spin-orbit coupling. This behavior becomes evident if one rewrites the Berry curvature as

[32] $\Omega_n(k) = i \sum_{m \neq n} \frac{\langle n|\frac{\partial H}{\partial k_1}|m\rangle\langle m|\frac{\partial H}{\partial k_2}|n\rangle - c.c.}{(E_n - E_m)^2}$. At the anticrossing points, the denominator becomes small, and the Berry curvature is enhanced. In our system, this corresponds to an enhancement of the spin Hall effect at the anticrossing points in figure 3(a).

Our results can also be understood as the consequence of a phase transition between two fundamental spin Hall states. As discussed by Zhu [36], band crossings and anticrossings correspond to phase transitions in a quantum system, and these transitions are accompanied by an enhancement of the Berry curvature. In our quantum wire system, this behavior is represented in figure 5, where we plot the spin-resolved magnitude squared of the wavefunctions representing subband **1+** in figure 5(a) and subband **2-** in figure 5(b), as a function of *k*. The transverse axis represents the position along the quantized axis of the wire. Yellow (light) corresponds to spin along +*z*, and red (dark) corresponds to spin along –*z*. We see that for low values of *k*, subband **1+** has one charge density peak, while subband **2-** has two. However, as they approach the anticrossing point, the wavefunctions hybridize and reach a state of similar symmetry with opposite spin displacements. For values of *k* above the anticrossing point, the subbands have actually swapped, as evidenced by the number of charge density peaks. This swapping is characteristic of a phase transition about the anticrossing point.

To calculate the Berry phase in a 2DEG in the presence of spin-orbit coupling, the most common approach is to perform a contour integral over momentum space. However, in a quantum wire system, a continuous integral over a 2D momentum vector is not possible due to the quantization of the system along one axis. We can overcome this difficulty if we rewrite the two-band Hamiltonian as

$$H(\theta) = \begin{bmatrix} \dfrac{\hbar^2 k^2}{2m^*} & \alpha_z e^{-i\theta}(ik_\parallel - k_\perp) \\ \alpha_z e^{i\theta}(-ik_\parallel - k_\perp) & \dfrac{\hbar^2 k^2}{2m^*} \end{bmatrix}, \quad (10)$$

where $\theta$ is the angle that the quantum wire makes with the [100] axis of the underlying crystal, $k_\parallel$ is the momentum operator along the length of the wire, and $k_\perp$ is the momentum operator in the transverse direction. This Hamiltonian can be quantized and applied to the Schrödinger equation as before, and the Berry phase of each subband can be calculated numerically as an integral over $\theta$, $\gamma_B^n = \int_0^{2\pi} \langle \phi_n | \dfrac{\partial}{\partial \theta} | \phi_n \rangle$. We find that when a particular subband in figure 3(a) has a positive slope, its Berry phase is given by $+\pi$, and when it has a negative slope its Berry phase is given by $-\pi$. Therefore, the anticrossing points symbolize a $2\pi$ transition in the Berry phase of the band, and its derivative with respect to *k*, the Berry curvature, exhibits a local maximum at these points. This differs from the infinite 2D case, where Shen [31] found that the Berry phase of each spin-split subband is equal to $\pi$. In (1), we see that the Rashba Hamiltonian depends not only on momentum, but also on spin. Therefore, it is possible to calculate the Berry phase of a particular subband as a contour integral over spin space. If we limit ourselves to spin vectors that lie in the 2D plane, and assume $\theta = 0$ in (10), then we can apply a rotator about the $z$-axis [37], $R_z(\omega) = \begin{bmatrix} e^{-i\omega/2} & 0 \\ 0 & e^{i\omega/2} \end{bmatrix}$, to the spin operator in the Rashba Hamiltonian. Doing so yields a Hamiltonian with the same form as in (10), except that $\theta$ is replaced by $\omega/2$. Therefore, the same results are obtained if our contour integral in spin space circles twice about the origin.

In summary, we have investigated the strength of the spin Hall effect in quantum wires as a function of electron density, spin-orbit coupling strength, and wire width. We found that maxima in the spin displacement are due to the anticrossing of adjacent subbands in momentum space. This led us to consider two approaches for optimizing the spin Hall effect in a quantum wire, by using a QPC to filter out all but the lowest subbands, or by using electrostatic top and back gates to reduce the density of electrons in the wire. We also discovered that the wire width that optimizes the spin Hall effect appears to be independent of disorder. The maxima in the spin displacement can be viewed as the result of a quantum phase transition between two fundamental spin Hall states, which manifests itself as a local maximum in the Berry curvature and subsequently in the spin displacement.

This document is the unedited author's version of a work that has been published in *J. Phys.: Condens. Matter* **21**, 055502 (2009). To access the final edited and published work see dx.doi.org/10.1088/0953-8984/21/5/055502.

AWC acknowledges the support of the DOE Computational Science Graduate Fellowship; grant number DE-FG02-97ER25308. Thanks also go out to Ethan Coxsey for his help in running simulations.

Figure 1. Spin displacement as a function of wire width assuming $n_{2D} = 4 \times 10^{11}$ cm$^{-2}$, considering (a) all occupied subbands, and (b) only the lowest pair of occupied subbands.

Figure 2. Spin displacement as a function of wire width assuming $\alpha_z = 20$ meV-nm, considering (a) all occupied subbands, and (b) only the lowest pair of occupied subbands.

Figure 3. (a) Electron dispersion relation of a 100 nm wire with a spin-orbit coupling strength of $\alpha_z = 20$ meV-nm. (b) Spin displacement as a function of $k$ for each of the subbands plotted in figure 3(a).

Figure 4. Spin displacement, in the presence of disorder, as a function of wire width assuming $n_{2D} = 4 \times 10^{11}$ cm$^{-2}$ and $\alpha_z = 20$ meV-nm, considering only the lowest pair of occupied subbands.

Figure 5. Magnitude squared, as a function of $k$, of the spin-resolved wave function associated with (a) subband **1+** and (b) subband **2-** in figure 3. Yellow (light) corresponds to spin along $+z$, while red (dark) corresponds to spin along $-z$.

Figure 1 Cummings *et al.*

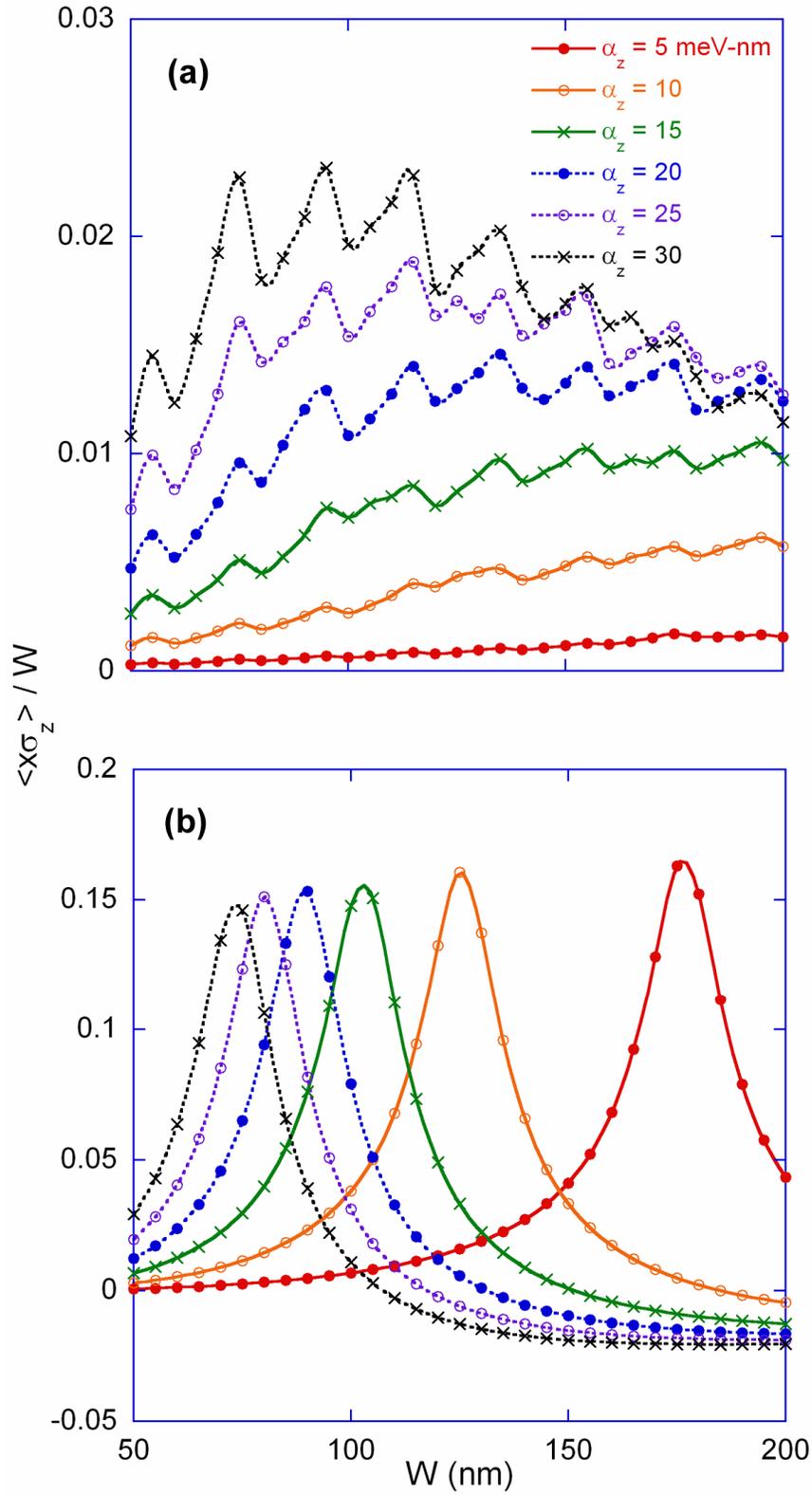

Figure 2 Cummings *et al.*

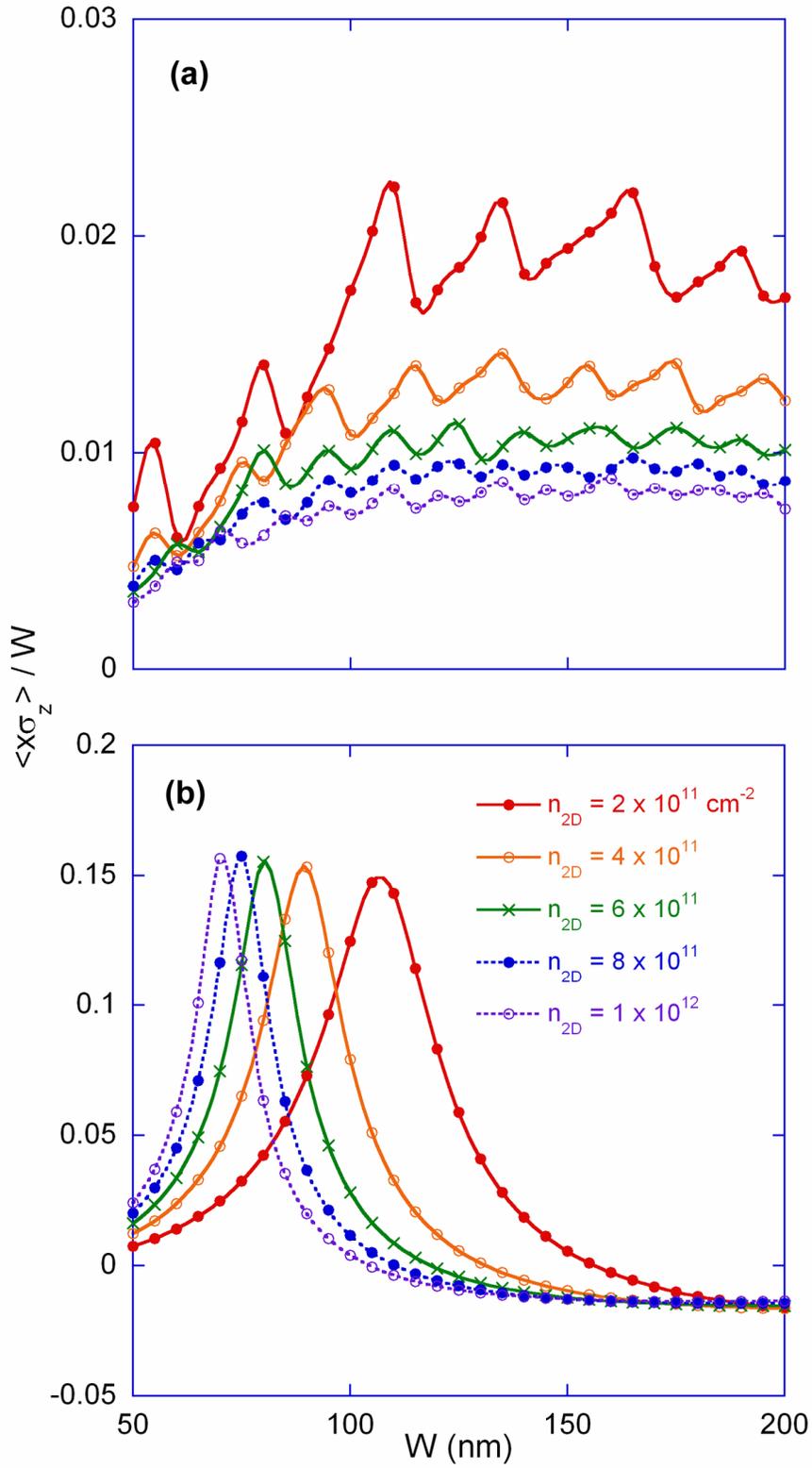

Figure 3 Cummings *et al.*

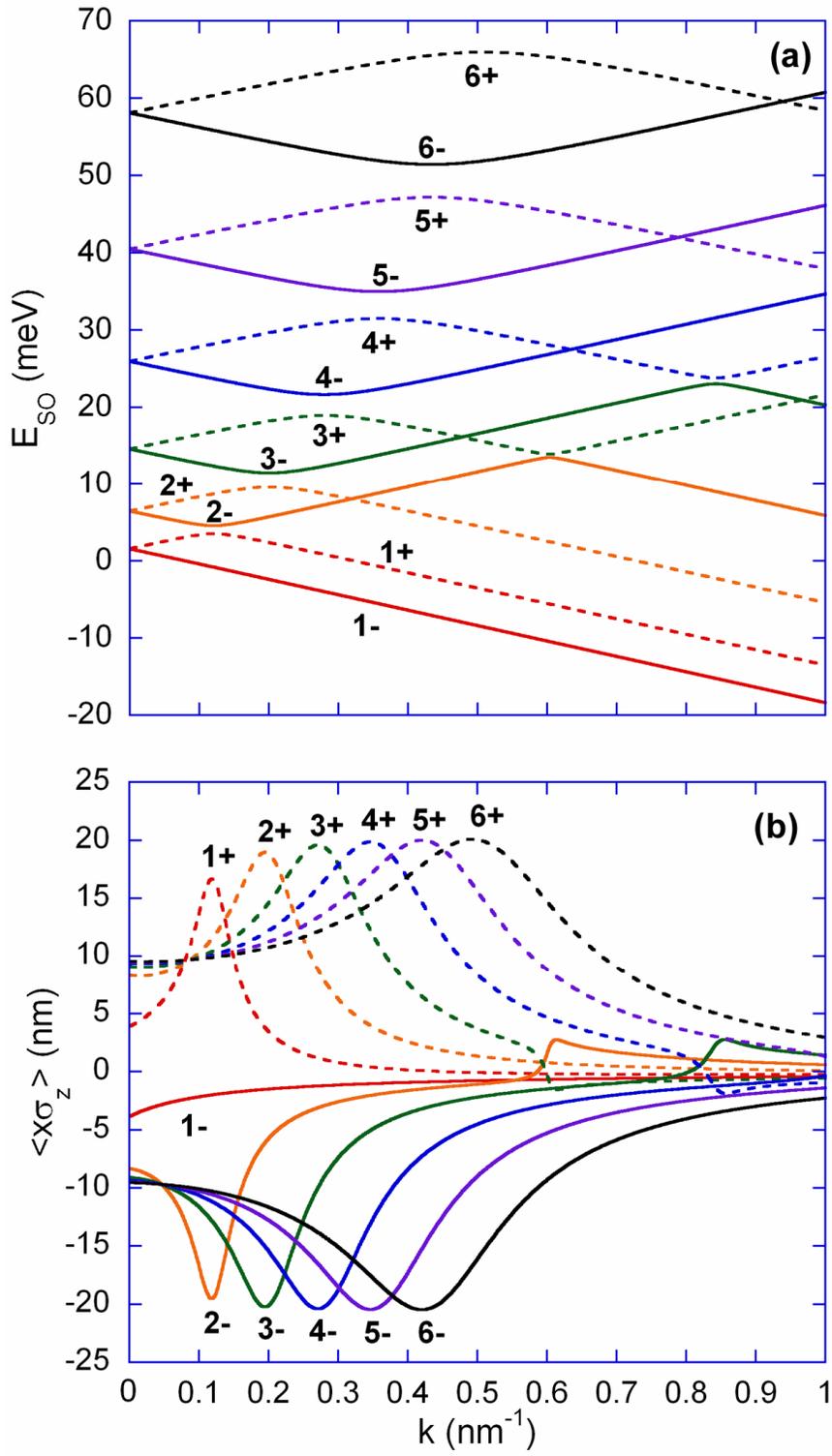

Figure 4 Cummings *et al.*

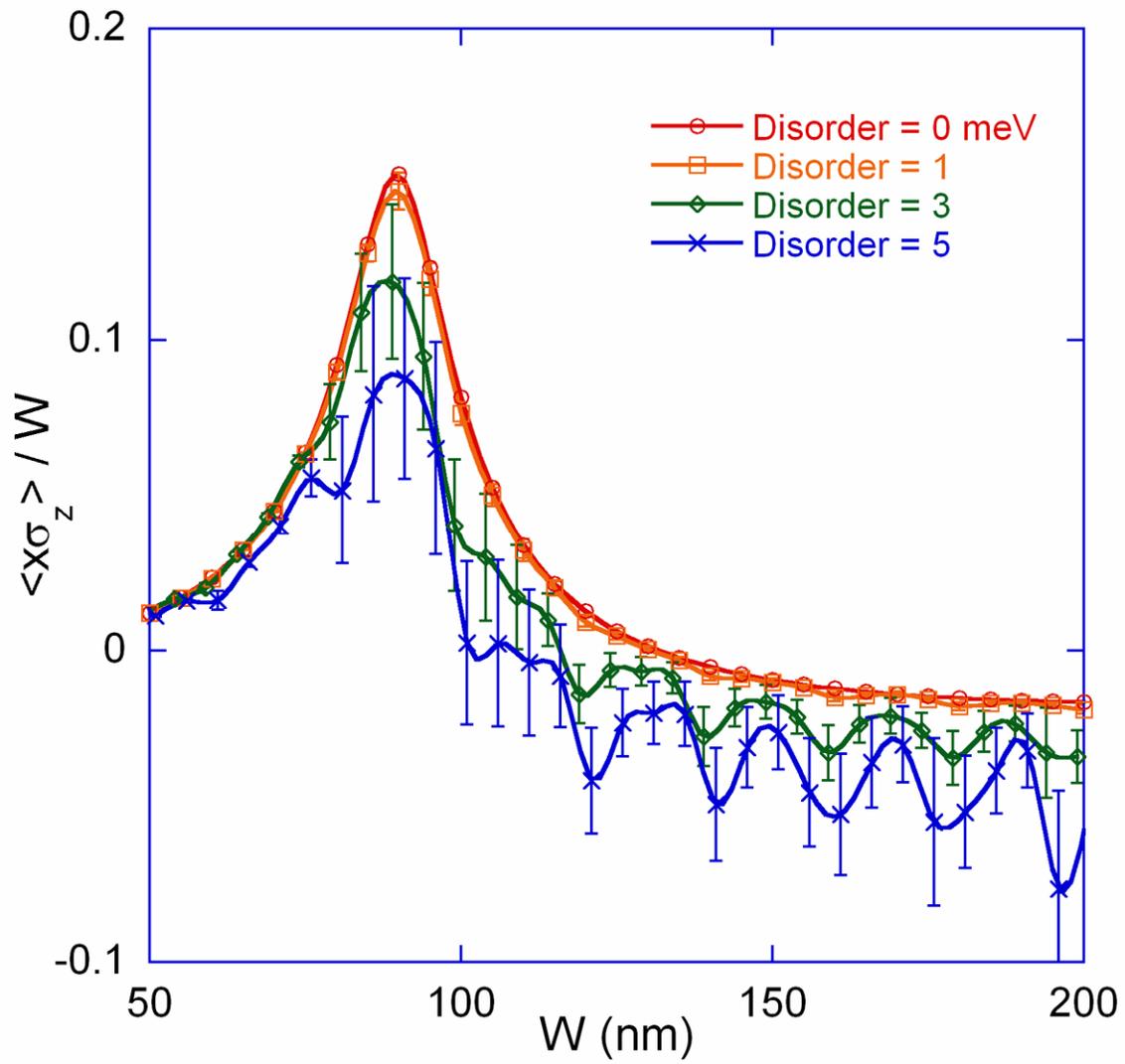

Figure 5 Cummings *et al.*

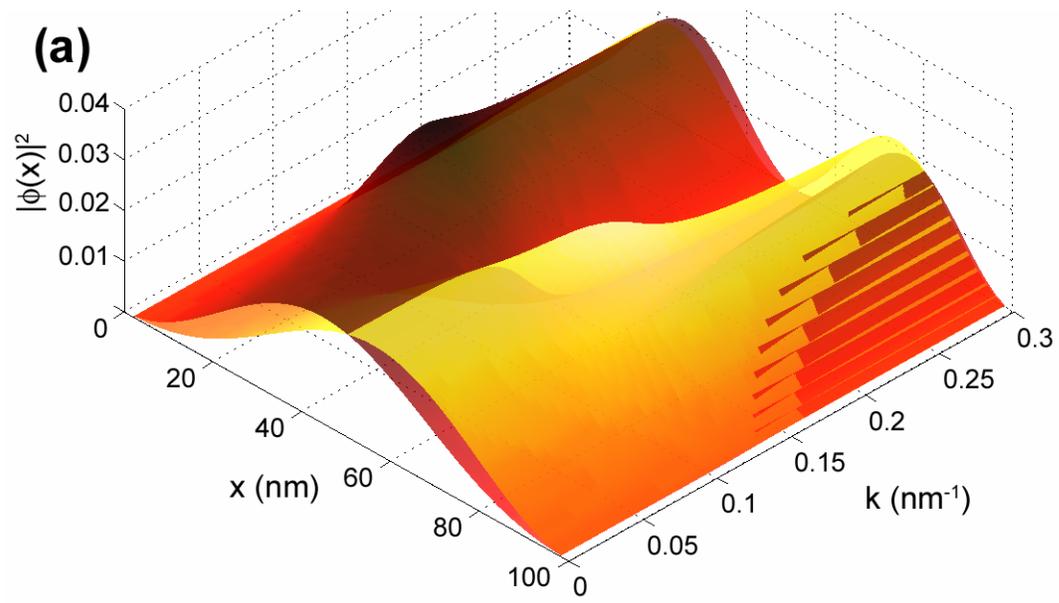

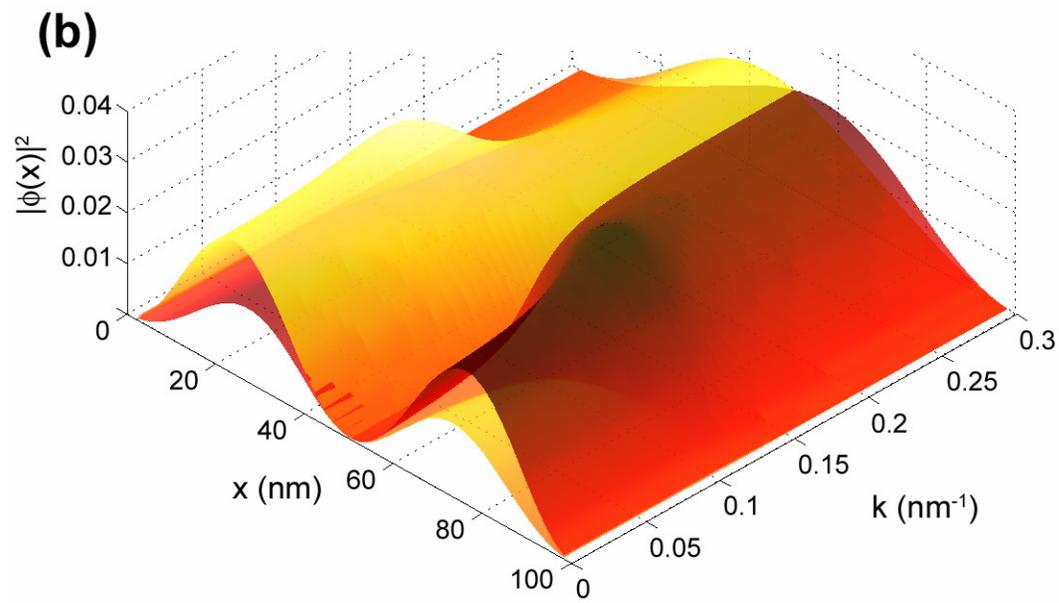